\documentclass[aps,prl,reprint,groupedaddress,showpacs]{revtex4-1}
\usepackage{amsmath}
\usepackage{amsfonts}
\usepackage{amssymb}
\usepackage{graphicx}	

\begin{document}

\title{Classical bifurcation at the transition from Rabi to Josephson dynamics}
\author{Tilman Zibold, Eike Nicklas, Christian Gross, Markus K. Oberthaler}
\affiliation{Kirchhoff Institut f\"ur Physik, Universit\"at Heidelberg, Im Neuenheimer Feld 227, D-69120, Germany}
\email[]{bifurcation@matterwave.de}
\date{\today}
\bibliographystyle{apsrev4-1}
\begin{abstract}
We report on the experimental realization of an internal bosonic Josephson junction in a Rubidium spinor Bose-Einstein condensate. The measurement of the full time dynamics in phase space allows the characterization of the theoretically predicted $\pi$-phase modes and quantitatively confirms analytical predictions, revealing a classical bifurcation. Our results suggest that this system is a model system which can be tuned from classical to the quantum regime and thus is an important step towards the experimental investigation of entanglement generation close to critical points.
\end{abstract}


\maketitle

Bifurcation occurs when a small smooth parameter change in a dynamical system leads to a sudden qualitative or topological change in its behavior. In classical nonlinear systems bifurcations are frequently encountered and are strongly related to critical phenomena and chaotic behavior \cite{Gutzwiller}. This relation is less obvious in the quantum regime due to the intrinsic uncertainty of the quantum states. However, macroscopic quantum systems exist which can be well described by classical theories exhibiting bifurcation phenomena \cite{Jessen,Hines,Courtney,Aubry,Siddiqi}. It has been theoretically shown that such a bifurcation can be used for the creation of highly entangled and nontrivial quantum states \cite{Micheli,Hines,Xie,Chuchem}. An exemplary system with these characteristics is the bosonic Josephson Junction \cite{Javanainen,Smerzi,Zapata,Kellman,Theocharis} which has so far been observed in weakly linked reservoirs of Helium \cite{Helium3,*Helium4} and Bose-Einstein condensates \cite{Cataliotti,Albiez,Steinhauer}. 

We report on the realization of a Josephson Junction in a Bose-Einstein condensate with internal i.e. spin degrees of freedom \cite{Williams} allowing the access of parameter regimes around the bifurcation point which have not been experimentally addressable yet. Since the experimental control of the tunneling coupling is realized via electromagnetic radiation the well developed techniques of precision spectroscopy can be employed to map out the full phase space i.e. dynamics of canonical conjugate variables, with high accuracy.

\begin{figure}[ht!]
\includegraphics[width=86mm]{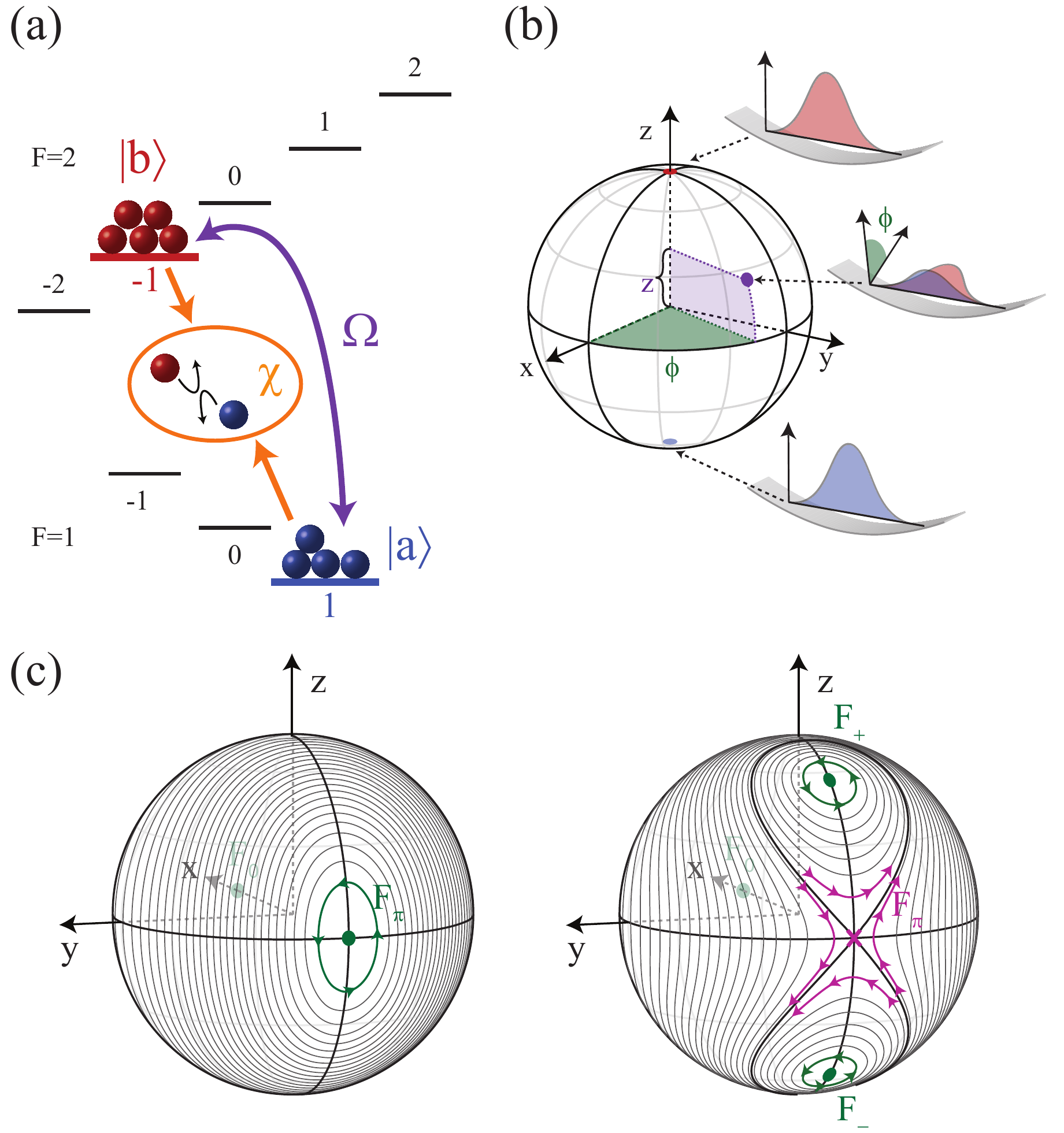}
\caption{\label{Fig1}(color online) Interacting many particle system as a model system for bifurcation physics. (a) ${}^\text{87}\text{Rb}$ offers two hyperfine states $|a\rangle$ (blue), $|b\rangle$ (red) which are linearly coupled via a two photon transition with Rabi frequency $\Omega$ and which allow for adjusting the inter particle interaction $\chi$ via a Feshbach resonance. (b) The many particle state is represented on a generalized Bloch sphere and its uncertainty area for our experimental parameters is shown revealing that a mean field description is adequate. Points on the sphere represent population difference $z$ (z-direction) and relative phase $\phi$ between the two internal states in the same spatial mode. (c) Trajectories on the Bloch sphere below and above the bifurcation value of the ratio $\Lambda=\chi N/\Omega$. The typical supercritical pitchfork bifurcation scenario occurs i.e. a stable fixed point bifurcates in two new stable fixed points while the original becomes unstable. The arrows indicate the direction of flow close to these points.}
\end{figure}
An internal Josephson junction is realized by N particles coherently distributed between two internal states $|a\rangle$ and $|b\rangle$. These states are linearly coupled with resonant two-photon radiofrequency-microwave radiation and experience coherent nonlinear interaction due to s-wave scattering between the atoms (see Fig. \ref{Fig1}). Assuming that both states are in the same spatial mode the dynamics is well described in the N particle two mode model with the Hamiltonian $H=\chi \hat{J}_z^2-\Omega \hat{J}_x$, where $\hat{\vec{J}}$ is the Schwinger pseudo spin representation of the $N$ atom system. $\hat{J}_z$ describes quantum mechanically the population difference between the two modes and $\hat{J}_x$ and $\hat{J}_y$ are corresponding coherences. Since the time evolution is given only by rotations in configuration space with the total number of particles conserved the dynamics can be visualized on a generalized Bloch sphere \cite{Gilmore} (see Fig. \ref{Fig1}b). The parameters $\chi$ and $\Omega$ describe the nonlinearity due to atom-atom interaction and the linear coupling strength respectively. It is interesting to note that this many particle Hamiltonian is a special case of the more general Lipkin-Meshkov-Glick model \cite{Lipkin} developed as a model system for theoretical studies in the context of nuclear physics.

In our experiments we investigate the dynamics of a macroscopic number of particles ($N=500$) and thus a mean field description is justified. This becomes obvious by comparing the uncertainty area of a coherent spin state -- on the order of $1/N$ -- to the surface of the Bloch sphere which is normalized to $4\pi$ (Fig. \ref{Fig1}b). The corresponding classical Hamiltonian is obtained by substituting the quantum mechanical operators by complex numbers such that  $H=\frac{\Lambda}{2} z^2-\sqrt{1-z^2}\cos{\phi}$ where $z=\frac{n_a-n_b}{N}$ is the normalized population difference and $\phi$ corresponds to the relative phase between the two internal states. The system parameters have been absorbed into the single parameter $\Lambda=\frac{\chi N}{\Omega}$ \cite{Raghavan}.

The equations of motion are given by:
\begin{eqnarray*}
\dot{z}(t)&=&-\sqrt{1-z^2(t)} \sin {\phi(t)} \\
\dot{\phi}(t)&=&\Lambda z(t)+\frac{z(t)}{\sqrt{1-z^2(t)}}\cos{\phi(t)}.
\end{eqnarray*}
Depending on the experimentally tunable parameter $\Lambda$, this system leads to qualitatively different dynamical behavior i.e. Rabi versus Josephson dynamics \cite{Leggett}. This becomes obvious by a classical fixed point analysis ($\dot{z}=0,\dot{\phi}=0$) which reveals the underlying topological change of phase space. For $\Lambda<1$, the Rabi regime, the linear coupling is governing the time evolution and two fixed points $F_0=(z,\phi)=(0,0)$ and $F_\pi=(0,\pi)$ characterize the dynamics. The trajectories are indicated by the solid lines in Fig. \ref{Fig1}c. For vanishing interaction between the particles, $\Lambda=0$, this corresponds to resonant Rabi oscillations of N independent particles. The situation changes drastically for $\Lambda>1$ since the $F_\pi$ fixed point undergoes a supercritical pitchfork bifurcation implying that $F_\pi$ becomes unstable while two new stable fixed points $F_{\pm}=(\pm \sqrt{1-\frac{1}{\Lambda^2}},\pi)$ are formed (Fig. \ref{Fig1}c).  For our system this implies that a single trajectory around $F_\pi$ splits up in two distinct trajectories around the new fixed points $F_\pm$ which are delimited by a separatrix.

\begin{figure}[t,floatfix]
\includegraphics[width=86mm]{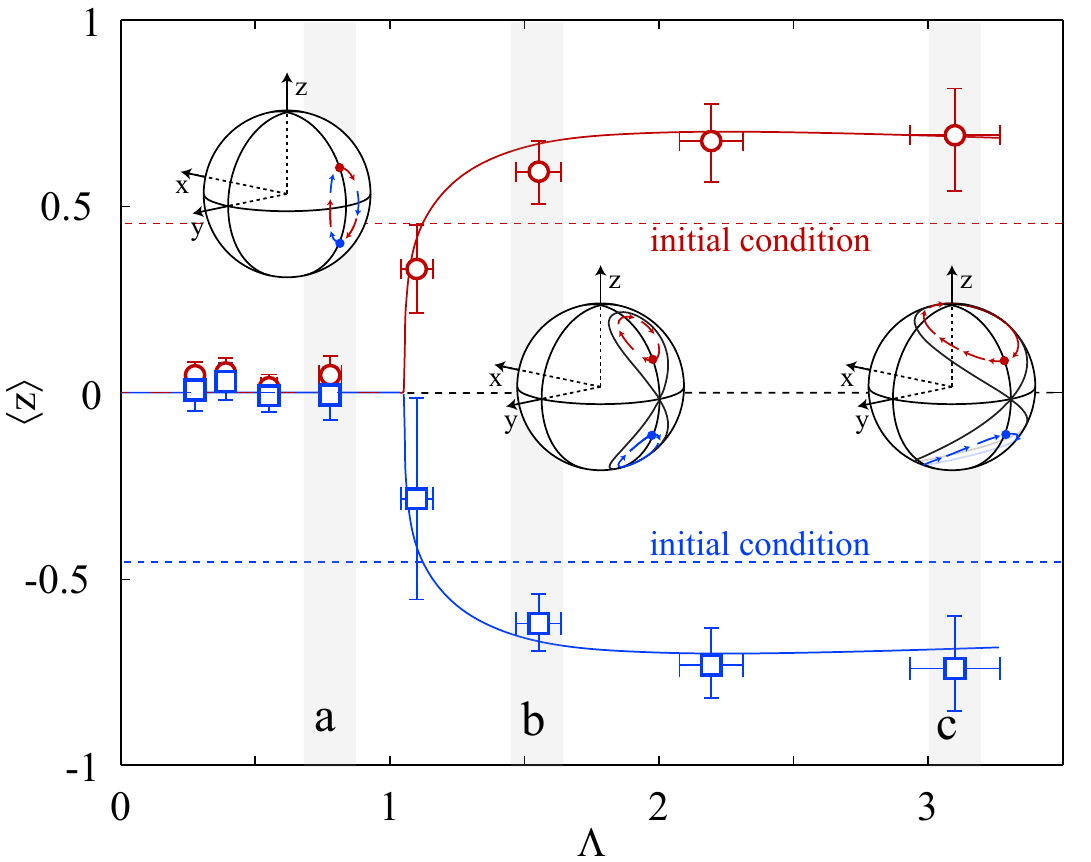}
\caption{\label{Fig2}(color online) Direct observation of the symmetry breaking in the dynamics due to the bifurcation. Two initial states symmetric in the upper and lower hemisphere (see inset) lead to qualitatively different dynamics in the Rabi and Josephson regime respectively. In the Rabi regime both initial states share the same trajectory around the stable fixed point $F_\pi$ and the temporal mean imbalance vanishes in both cases. By increasing $\Lambda$ exceeding the critical value a separatrix is formed and the chosen initial preparations lead to two distinct trajectories separated by this separatrix. The dynamical modes are characterized by a non-vanishing mean population imbalance. The solid line represents the theoretical prediction.}
\end{figure}

For a quantitative experimental study of the bifurcation phenomenon we study the temporal mean imbalance for two fixed initial preparations. In the Rabi regime ($\Lambda<1$) initial preparations with $\phi=\pi$ and $z=\pm z_0$ corresponding to points north/south of the equator (see inset Fig. \ref{Fig2}) lead to dynamics with a vanishing temporal mean population imbalance. This results from the fact that both preparations share the same trajectory i.e. no separatrix exists. This is distinct to the Josephson regime where initial preparations that are enclosed by the separatrix lead to different trajectories resulting in non vanishing mean imbalances. This is demonstrated quantitatively in Figure \ref{Fig2} where the resulting temporal mean imbalances for the initial preparation points $(\pm 0.454,\pi)$ are shown. The experimental data clearly reveals the topological change in the system's phase space. It is in quantitative agreement with the analytical predictions (solid lines) \cite{Raghavan} calculated using independently measured parameters (see \cite{EPAPS}).

\begin{figure}[t]
\includegraphics[width=86mm]{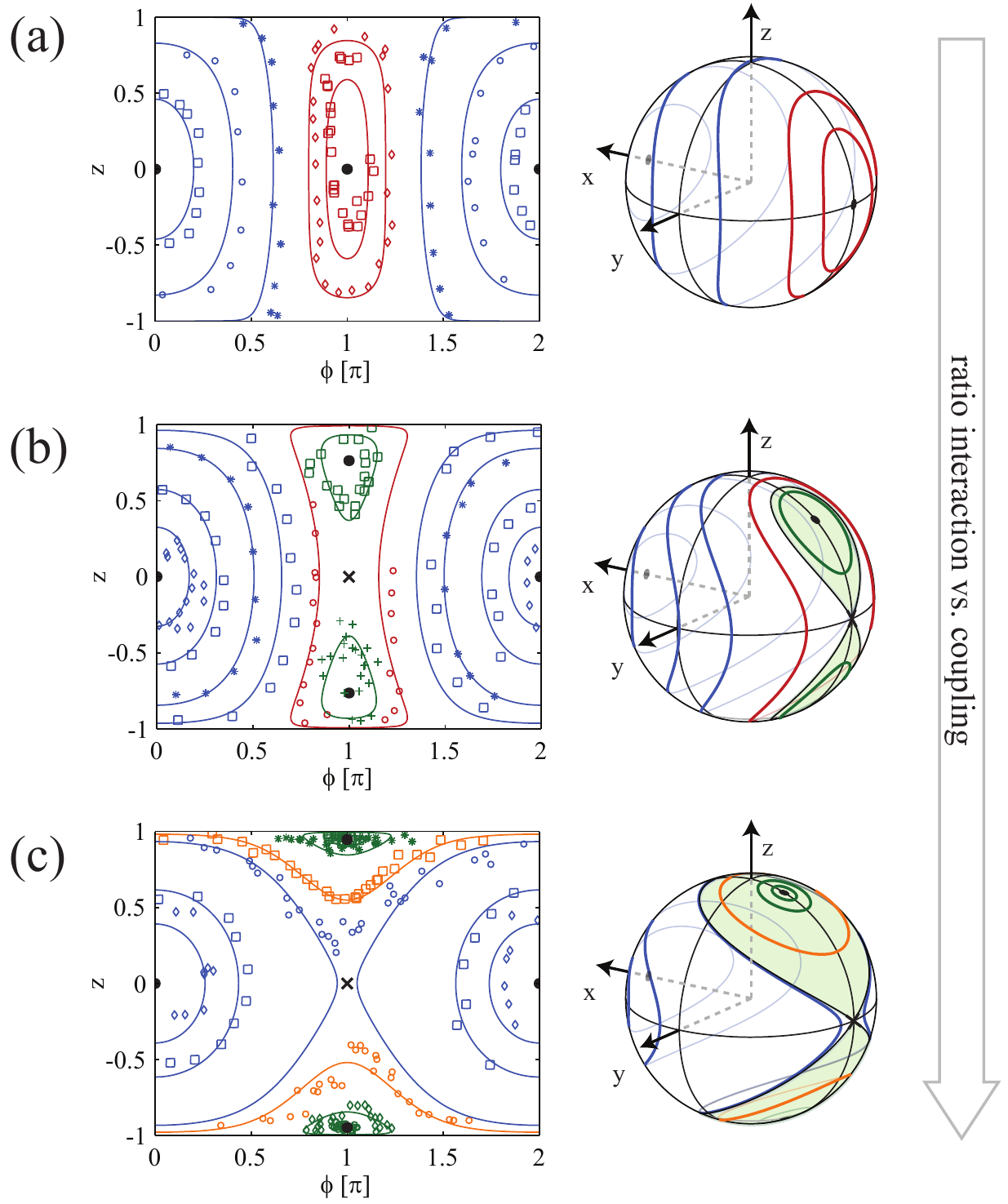}
\caption{\label{Fig3}(color online) Experimentally observed phase portraits of the dynamics showing all possible types of trajectories . The experimental data for three different $\Lambda$ are compared to the theoretical model with no free parameter (solid lines). The theoretical curves are shown additionally on the Bloch spheres. (a), Phase portrait in the Rabi regime for $\Lambda=0.78$. Plasma-  (blue) and $\pi$-oscillations (red) can be clearly identified.  The deformation relative to non-interacting Rabi-oscillations due to the interaction is clearly visible on the Bloch sphere. (b), The Josephson regime is entered by reducing the coupling strength $\Omega$ ($\Lambda=1.55$). Here the bifurcation leads to new stable solutions showing macroscopic quantum self trapping with mean phase $\pi$. This is verified by the experimental data (green squares and green crosses). The black solid line on the Bloch sphere corresponds to the separatrix defining the green shaded area of macroscopic quantum self trapping. (c), For $\Lambda>2$ the separatrix encloses the two poles and self trapped trajectories with running phase behavior are found experimentally for $\Lambda=3.1$ (orange squares and circles). These trajectories resemble the ac-Josephson effect found in superconducting Josephson-junctions.}
\end{figure}

To put this bifurcation measurement in a more general context we examine the whole phase portrait of the system for characteristic values of $\Lambda$ across the Rabi to Josephson transition. The nonlinear interaction $\chi$ is set by a Feshbach resonance at 9.1 G \cite{Widera,*Sengstock} and is kept constant for all experiments. Different regimes of $\Lambda$ are explored by changing the linear coupling strength $\Omega$ adjusted by the intensity of the radio frequency radiation. We check the resonant coupling condition by regular reference measurements \cite{EPAPS}. The measurement of the dynamics with shot noise limited precision is feasible in our experiment since we prepare the initial condition on the quantum mechanical uncertainty level i.e. coherent spin states \cite{Gross}. The initial state preparation is done in a two step process. The population imbalance $z(t=0)$ is controlled by the duration of a short two photon pulse applied to the particles in state $| a\rangle$. The dynamics is initiated by a non-adiabatic change of the radio frequency radiation phase of $\phi(t=0)=\phi_0$ and simultaneous attenuation of the radiation. After an evolution time $t$ either the population difference $z(t)$ of the final state is directly measured or the phase $\phi(t)$ is mapped onto an observable population imbalance by applying a short $\pi/2$-pulse before imaging. This last pulse is applied with two phases differing by $90^\circ$ to get a well defined phase measurement over the full interval $0$ to $2\pi$.  By repeated experiments we are able to measure both observables allowing the mapping out of the phase portrait.

Figure \ref{Fig3}a shows the result in the Rabi regime, where the linear coupling is dominating. The dynamics is characterized by two fixed points however the corresponding trajectories around these are differently distorted. In the literature the motion around $F_0$ is known as plasma oscillations already experimentally observed \cite{Albiez,Steinhauer} while the trajectories around $F_\pi$ known as $\pi$-oscillations, have not been demonstrated so far. By reducing the linear coupling the Josephson regime is entered. The transition is marked by the bifurcation of $F_\pi$ as seen in Figure \ref{Fig3}b and the dynamics corresponding to the new fixed points is known as macroscopic quantum self trapping. This describes the physical fact that the temporal mean population imbalance is non zero. These modes (green shaded area in Fig. \ref{Fig3}b and \ref{Fig3}c) are separated in phase space by the separatrix from the plasma and $\pi$-oscillations which are characterized by vanishing temporal mean $\langle z \rangle=0$. By further increasing $\Lambda$ the topology does not change anymore but the trajectories start to encircle the north/south poles of the sphere. This implies that the phase evolution runs without bound and connects the observed dynamics with the phenomenon of ac-Josephson effect found in superconductors \cite{Josephson}. It is important to note that the full analogy is not given due to the spherical topology arising from particle number conservation. In superconductors charge neutrality implies that the chemical potential difference is kept constant and thus the dynamics is rather constraint to a cylinder. However the analogy is given best for large $\Lambda$ and small absolute imbalance $|z|$ \cite{Radzihovsky}.  

\begin{figure}[t]
\includegraphics[width=86mm]{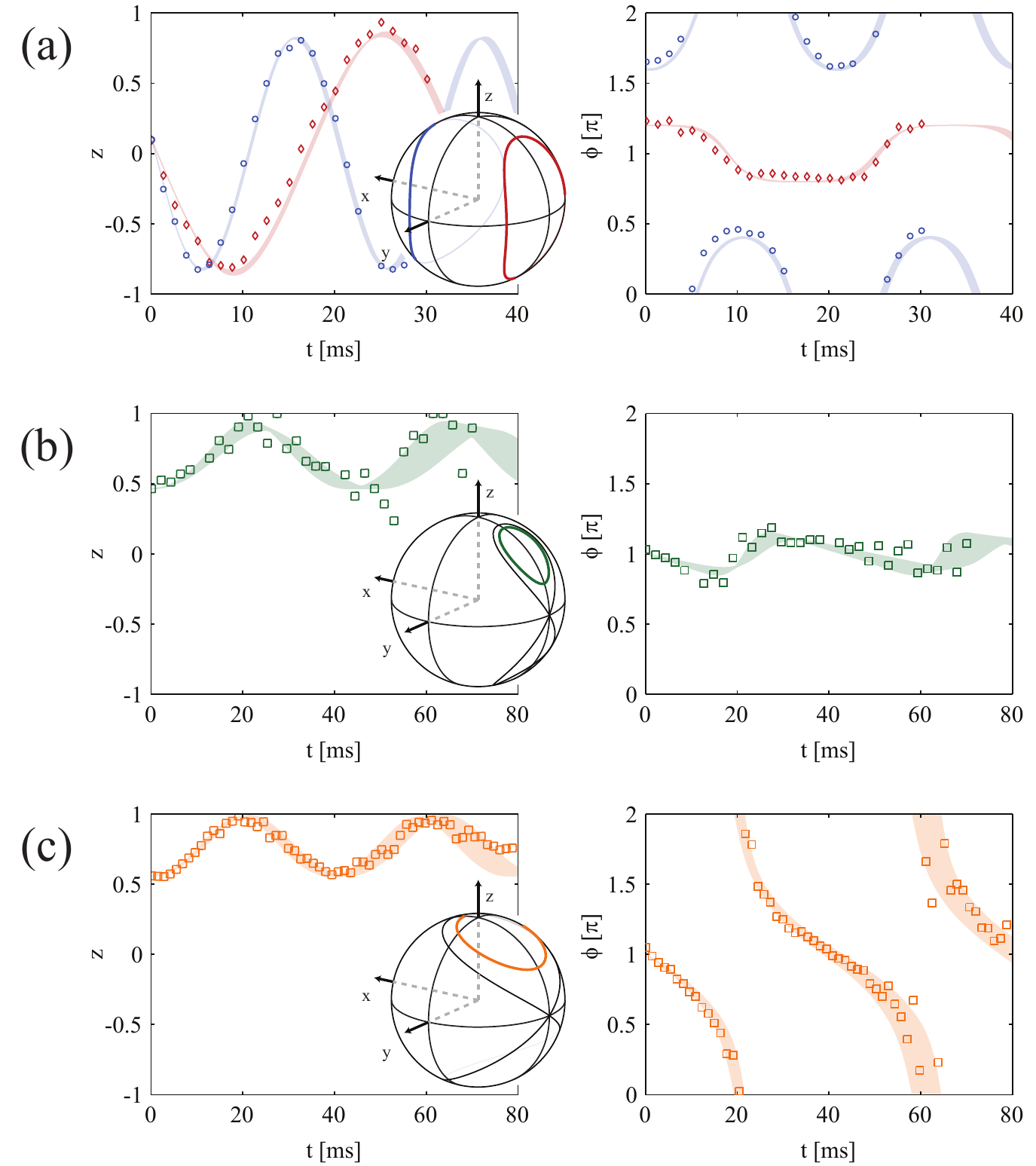}
\caption{\label{Fig4}(color online) Exemplary time dynamics of population imbalance and relative phase. (a) Plasma- (blue) and $\pi$-oscillations (red) in the Rabi regime ($\Lambda=0.78$). The dynamics around the two fixed points is qualitatively the same. However, the oscillations with mean phase $\pi$ are slower than their counterparts with vanishing mean phase. In the former case the atom interaction counteracts, whereas in the latter it assists the linear coupling . The shaded area corresponds to the theoretical prediction including the experimental uncertainty in $\Lambda$.  (b) Macroscopic quantum self trapping for $\Lambda=1.55$. Going beyond the critical value $\Lambda$ for bifurcation the Josephson regime is entered. Time traces show oscillations around the northern fixed point $F_+$ while the mean phase varies around $\pi$. (c) Running phase mode found at $\Lambda=3.1$. The system is prepared on a trajectory that encircles the north pole of the sphere. Dynamics in the population imbalance is similar to (b) but the phase runs without bound as it is the case for the ac-Josephson effect in superconductors.}
\end{figure}

The  full time dynamics i.e. particle number difference and phase difference, is depicted in Fig. \ref{Fig4}. The shaded areas represent the theoretical expectations without free parameters but including experimental uncertainties of 5\% in $\Lambda$.  In the limit of vanishing interaction ($\Lambda=0$) plasma and pi oscillations are not distinguishable.
For finite $\Lambda$ but still in the Rabi regime  (Fig. \ref{Fig4}a) the difference between plasma and $\pi$ oscillations manifests itself most pronouncedly in the modified oscillation frequency which is reduced in the case of $\pi$-oscillations and enhanced for plasma oscillations \cite{Raghavan}. Crossing the critical value of $\Lambda$ i.e. in the bifurcated regime, the new dynamical mode, known as macroscopic quantum self trapping is observed (Fig. \ref{Fig4}b and \ref{Fig4}c). In the Josephson regime, but close to the critical value ($1<\Lambda<2$) all self trapping modes have an oscillating phase difference (Fig. \ref{Fig4}b), while for $\Lambda>2$ also running phase self trapping modes exist (Fig. \ref{Fig4}c).

In this work we demonstrate the experimental realization of a quantum mechanical many particle system exhibiting a classical bifurcation. This opens up a new experimental route for generating non-trivial collective quantum states. It has been theoretically discussed that a general feature of the realized system is the potential for fast generation of macroscopic entanglement at the bifurcation point or close to the separatrix \cite{Micheli}. The demonstrated high level of experimental control together with the ability of precise measurement of conjugate collective variables \cite{Gross} makes our system a model system for these kinds of investigations and allows the generation as well as the study of nontrivial many particle quantum dynamics in the macroscopic regime. Furthermore, the possibility to scale the effective Planck's constant, i.e. the uncertainty area of a coherent spin state by adjusting the total atom number makes this a unique experimental system for studying of the cross over from classical and quantum physics \cite{Chuchem}.

\begin{acknowledgements}
We thank Helmut Strobel for technical help with calibration measurements. We gratefully acknowledge support from the Heidelberg Center of Quantum Dynamics, the Deutsche Forschungsgemeinschaft for support of the Forschergruppe 760, the German-Israeli Foundation, the ExtreMe Matter Institute and the European Commission FET open scheme project MIDAS. T.Z. and C.G. acknowlege support from the Landesgraduiertenf\"orderung Baden-W\"urttemberg.
\end{acknowledgements}


\begin{thebibliography}{30}%
\makeatletter
\providecommand \@ifxundefined [1]{%
 \@ifx{#1\undefined}
}%
\providecommand \@ifnum [1]{%
 \ifnum #1\expandafter \@firstoftwo
 \else \expandafter \@secondoftwo
 \fi
}%
\providecommand \@ifx [1]{%
 \ifx #1\expandafter \@firstoftwo
 \else \expandafter \@secondoftwo
 \fi
}%
\providecommand \natexlab [1]{#1}%
\providecommand \enquote  [1]{``#1''}%
\providecommand \bibnamefont  [1]{#1}%
\providecommand \bibfnamefont [1]{#1}%
\providecommand \citenamefont [1]{#1}%
\providecommand \href@noop [0]{\@secondoftwo}%
\providecommand \href [0]{\begingroup \@sanitize@url \@href}%
\providecommand \@href[1]{\@@startlink{#1}\@@href}%
\providecommand \@@href[1]{\endgroup#1\@@endlink}%
\providecommand \@sanitize@url [0]{\catcode `\\12\catcode `\$12\catcode
  `\&12\catcode `\#12\catcode `\^12\catcode `\_12\catcode `\%12\relax}%
\providecommand \@@startlink[1]{}%
\providecommand \@@endlink[0]{}%
\providecommand \url  [0]{\begingroup\@sanitize@url \@url }%
\providecommand \@url [1]{\endgroup\@href {#1}{\urlprefix }}%
\providecommand \urlprefix  [0]{URL }%
\providecommand \Eprint [0]{\href }%
\@ifxundefined \urlstyle {%
  \providecommand \doi  [0]{\begingroup \@sanitize@url \@doi}%
  \providecommand \@doi [1]{\endgroup \@@startlink {\doibase
  #1}doi:\discretionary {}{}{}#1\@@endlink }%
}{%
  \providecommand \doi  [0]{doi:\discretionary{}{}{}\begingroup
  \urlstyle{rm}\Url }%
}%
\providecommand \doibase [0]{http://dx.doi.org/}%
\providecommand \Doi [0]{\begingroup \@sanitize@url \@Doi }%
\providecommand \@Doi  [1]{\endgroup\@@startlink{\doibase#1}\@@Doi}%
\providecommand \@@Doi [1]{#1\@@endlink}%
\providecommand \selectlanguage [0]{\@gobble}%
\providecommand \bibinfo  [0]{\@secondoftwo}%
\providecommand \bibfield  [0]{\@secondoftwo}%
\providecommand \translation [1]{[#1]}%
\providecommand \BibitemOpen [0]{}%
\providecommand \bibitemStop [0]{}%
\providecommand \bibitemNoStop [0]{.\EOS\space}%
\providecommand \EOS [0]{\spacefactor3000\relax}%
\providecommand \BibitemShut  [1]{\csname bibitem#1\endcsname}%
\bibitem [{\citenamefont {Gutzwiller}(1991)}]{Gutzwiller}%
  \BibitemOpen
  \bibfield  {author} {\bibinfo {author} {\bibfnamefont {M.~C.}\ \bibnamefont
  {Gutzwiller}},\ }\href@noop {} {\emph {\bibinfo {title} {Chaos in Classical
  and Quantum Mechanics}}}\ (\bibinfo  {publisher} {Springer-Verlag, New
  York},\ \bibinfo {year} {1991})\BibitemShut {NoStop}%
\bibitem [{\citenamefont {Chaudhury}\ \emph {et~al.}(2009)\citenamefont
  {Chaudhury}, \citenamefont {Smith}, \citenamefont {Anderson}, \citenamefont
  {Ghose},\ and\ \citenamefont {Jessen}}]{Jessen}%
  \BibitemOpen
  \bibfield  {author} {\bibinfo {author} {\bibfnamefont {S.}~\bibnamefont
  {Chaudhury}}, \bibinfo {author} {\bibfnamefont {A.}~\bibnamefont {Smith}},
  \bibinfo {author} {\bibfnamefont {B.~E.}\ \bibnamefont {Anderson}}, \bibinfo
  {author} {\bibfnamefont {S.}~\bibnamefont {Ghose}}, \ and\ \bibinfo {author}
  {\bibfnamefont {P.~S.}\ \bibnamefont {Jessen}},\ }\href
  {http://dx.doi.org/10.1038/nature08396} {\bibfield  {journal} {\bibinfo
  {journal} {Nature (London)},\ }\textbf {\bibinfo {volume} {461}},\ \bibinfo
  {pages} {768} (\bibinfo {year} {2009})}\BibitemShut {NoStop}%
\bibitem [{\citenamefont {Hines}\ \emph {et~al.}(2005)\citenamefont {Hines},
  \citenamefont {McKenzie},\ and\ \citenamefont {Milburn}}]{Hines}%
  \BibitemOpen
  \bibfield  {author} {\bibinfo {author} {\bibfnamefont {A.~P.}\ \bibnamefont
  {Hines}}, \bibinfo {author} {\bibfnamefont {R.~H.}\ \bibnamefont {McKenzie}},
  \ and\ \bibinfo {author} {\bibfnamefont {G.~J.}\ \bibnamefont {Milburn}},\
  }\Doi {10.1103/PhysRevA.71.042303} {\bibfield  {journal} {\bibinfo  {journal}
  {Phys. Rev. A},\ }\textbf {\bibinfo {volume} {71}},\ \bibinfo {pages}
  {042303} (\bibinfo {year} {2005})}\BibitemShut {NoStop}%
\bibitem [{\citenamefont {Courtney}\ \emph {et~al.}(1995)\citenamefont
  {Courtney}, \citenamefont {Jiao}, \citenamefont {Spellmeyer}, \citenamefont
  {Kleppner}, \citenamefont {Gao},\ and\ \citenamefont {Delos}}]{Courtney}%
  \BibitemOpen
  \bibfield  {author} {\bibinfo {author} {\bibfnamefont {M.}~\bibnamefont
  {Courtney}}, \bibinfo {author} {\bibfnamefont {H.}~\bibnamefont {Jiao}},
  \bibinfo {author} {\bibfnamefont {N.}~\bibnamefont {Spellmeyer}}, \bibinfo
  {author} {\bibfnamefont {D.}~\bibnamefont {Kleppner}}, \bibinfo {author}
  {\bibfnamefont {J.}~\bibnamefont {Gao}}, \ and\ \bibinfo {author}
  {\bibfnamefont {J.~B.}\ \bibnamefont {Delos}},\ }\Doi
  {10.1103/PhysRevLett.74.1538} {\bibfield  {journal} {\bibinfo  {journal}
  {Phys. Rev. Lett.},\ }\textbf {\bibinfo {volume} {74}},\ \bibinfo {pages}
  {1538} (\bibinfo {year} {1995})}\BibitemShut {NoStop}%
\bibitem [{\citenamefont {Aubry}\ \emph {et~al.}(1996)\citenamefont {Aubry},
  \citenamefont {Flach}, \citenamefont {Kladko},\ and\ \citenamefont
  {Olbrich}}]{Aubry}%
  \BibitemOpen
  \bibfield  {author} {\bibinfo {author} {\bibfnamefont {S.}~\bibnamefont
  {Aubry}}, \bibinfo {author} {\bibfnamefont {S.}~\bibnamefont {Flach}},
  \bibinfo {author} {\bibfnamefont {K.}~\bibnamefont {Kladko}}, \ and\ \bibinfo
  {author} {\bibfnamefont {E.}~\bibnamefont {Olbrich}},\ }\href@noop {}
  {\bibfield  {journal} {\bibinfo  {journal} {Phys. Rev. Lett.},\ }\textbf
  {\bibinfo {volume} {76}},\ \bibinfo {pages} {1607} (\bibinfo {year}
  {1996})}\BibitemShut {NoStop}%
\bibitem [{\citenamefont {Siddiqi}\ \emph {et~al.}(2005)\citenamefont
  {Siddiqi}, \citenamefont {Vijay}, \citenamefont {Pierre}, \citenamefont
  {Wilson}, \citenamefont {Frunzio}, \citenamefont {Metcalfe}, \citenamefont
  {Rigetti}, \citenamefont {Schoelkopf}, \citenamefont {Devoret}, \citenamefont
  {Vion},\ and\ \citenamefont {Esteve}}]{Siddiqi}%
  \BibitemOpen
  \bibfield  {author} {\bibinfo {author} {\bibfnamefont {I.}~\bibnamefont
  {Siddiqi}}, \bibinfo {author} {\bibfnamefont {R.}~\bibnamefont {Vijay}},
  \bibinfo {author} {\bibfnamefont {F.}~\bibnamefont {Pierre}}, \bibinfo
  {author} {\bibfnamefont {C.~M.}\ \bibnamefont {Wilson}}, \bibinfo {author}
  {\bibfnamefont {L.}~\bibnamefont {Frunzio}}, \bibinfo {author} {\bibfnamefont
  {M.}~\bibnamefont {Metcalfe}}, \bibinfo {author} {\bibfnamefont
  {C.}~\bibnamefont {Rigetti}}, \bibinfo {author} {\bibfnamefont {R.~J.}\
  \bibnamefont {Schoelkopf}}, \bibinfo {author} {\bibfnamefont {M.~H.}\
  \bibnamefont {Devoret}}, \bibinfo {author} {\bibfnamefont {D.}~\bibnamefont
  {Vion}}, \ and\ \bibinfo {author} {\bibfnamefont {D.}~\bibnamefont
  {Esteve}},\ }\Doi {10.1103/PhysRevLett.94.027005} {\bibfield  {journal}
  {\bibinfo  {journal} {Phys. Rev. Lett.},\ }\textbf {\bibinfo {volume} {94}},\
  \bibinfo {pages} {027005} (\bibinfo {year} {2005})}\BibitemShut {NoStop}%
\bibitem [{\citenamefont {Micheli}\ \emph {et~al.}(2003)\citenamefont
  {Micheli}, \citenamefont {Jaksch}, \citenamefont {Cirac},\ and\ \citenamefont
  {Zoller}}]{Micheli}%
  \BibitemOpen
  \bibfield  {author} {\bibinfo {author} {\bibfnamefont {A.}~\bibnamefont
  {Micheli}}, \bibinfo {author} {\bibfnamefont {D.}~\bibnamefont {Jaksch}},
  \bibinfo {author} {\bibfnamefont {J.}~\bibnamefont {Cirac}}, \ and\ \bibinfo
  {author} {\bibfnamefont {P.}~\bibnamefont {Zoller}},\ }\Doi
  {10.1103/PhysRevA.67.013607} {\bibfield  {journal} {\bibinfo  {journal}
  {Phys. Rev. A},\ }\textbf {\bibinfo {volume} {67}},\ \bibinfo {pages}
  {013607} (\bibinfo {year} {2003})}\BibitemShut {NoStop}%
\bibitem [{\citenamefont {Xie}\ and\ \citenamefont {Hai}(2006)}]{Xie}%
  \BibitemOpen
  \bibfield  {author} {\bibinfo {author} {\bibfnamefont {Q.}~\bibnamefont
  {Xie}}\ and\ \bibinfo {author} {\bibfnamefont {W.}~\bibnamefont {Hai}},\
  }\Doi {10.1140/epjd/e2006-00103-6} {\bibfield  {journal} {\bibinfo  {journal}
  {Eur. Phys. J. D},\ }\textbf {\bibinfo {volume} {39}},\ \bibinfo {pages}
  {277} (\bibinfo {year} {2006})}\BibitemShut {NoStop}%
\bibitem [{\citenamefont {Chuchem}\ \emph {et~al.}()\citenamefont {Chuchem},
  \citenamefont {Smith-Mannschott}, \citenamefont {Hiller}, \citenamefont
  {Kottos}, \citenamefont {Vardi},\ and\ \citenamefont {Cohen}}]{Chuchem}%
  \BibitemOpen
  \bibfield  {author} {\bibinfo {author} {\bibfnamefont {M.}~\bibnamefont
  {Chuchem}}, \bibinfo {author} {\bibfnamefont {K.}~\bibnamefont
  {Smith-Mannschott}}, \bibinfo {author} {\bibfnamefont {M.}~\bibnamefont
  {Hiller}}, \bibinfo {author} {\bibfnamefont {T.}~\bibnamefont {Kottos}},
  \bibinfo {author} {\bibfnamefont {A.}~\bibnamefont {Vardi}}, \ and\ \bibinfo
  {author} {\bibfnamefont {D.}~\bibnamefont {Cohen}},\ }\href@noop {} {}\Eprint
  {http://arxiv.org/abs/1001.2120} {arXiv:1001.2120} \BibitemShut {NoStop}%
\bibitem [{\citenamefont {Javanainen}(1986)}]{Javanainen}%
  \BibitemOpen
  \bibfield  {author} {\bibinfo {author} {\bibfnamefont {J.}~\bibnamefont
  {Javanainen}},\ }\Doi {10.1103/PhysRevLett.57.3164} {\bibfield  {journal}
  {\bibinfo  {journal} {Phys. Rev. Lett.},\ }\textbf {\bibinfo {volume} {57}},\
  \bibinfo {pages} {3164} (\bibinfo {year} {1986})}\BibitemShut {NoStop}%
\bibitem [{\citenamefont {Smerzi}\ \emph {et~al.}(1997)\citenamefont {Smerzi},
  \citenamefont {Fantoni}, \citenamefont {Giovanazzi},\ and\ \citenamefont
  {Shenoy}}]{Smerzi}%
  \BibitemOpen
  \bibfield  {author} {\bibinfo {author} {\bibfnamefont {A.}~\bibnamefont
  {Smerzi}}, \bibinfo {author} {\bibfnamefont {S.}~\bibnamefont {Fantoni}},
  \bibinfo {author} {\bibfnamefont {S.}~\bibnamefont {Giovanazzi}}, \ and\
  \bibinfo {author} {\bibfnamefont {S.}~\bibnamefont {Shenoy}},\ }\href@noop {}
  {\bibfield  {journal} {\bibinfo  {journal} {Phys. Rev. Lett.},\ }\textbf
  {\bibinfo {volume} {79}},\ \bibinfo {pages} {4950} (\bibinfo {year}
  {1997})}\BibitemShut {NoStop}%
\bibitem [{\citenamefont {Zapata}\ \emph {et~al.}(1998)\citenamefont {Zapata},
  \citenamefont {Sols},\ and\ \citenamefont {Leggett}}]{Zapata}%
  \BibitemOpen
  \bibfield  {author} {\bibinfo {author} {\bibfnamefont {I.}~\bibnamefont
  {Zapata}}, \bibinfo {author} {\bibfnamefont {F.}~\bibnamefont {Sols}}, \ and\
  \bibinfo {author} {\bibfnamefont {A.}~\bibnamefont {Leggett}},\ }\href@noop
  {} {\bibfield  {journal} {\bibinfo  {journal} {Phys. Rev. A},\ }\textbf
  {\bibinfo {volume} {57}},\ \bibinfo {pages} {R28} (\bibinfo {year}
  {1998})}\BibitemShut {NoStop}%
\bibitem [{\citenamefont {Kellman}\ and\ \citenamefont {Tyng}(2002)}]{Kellman}%
  \BibitemOpen
  \bibfield  {author} {\bibinfo {author} {\bibfnamefont {M.~E.}\ \bibnamefont
  {Kellman}}\ and\ \bibinfo {author} {\bibfnamefont {V.}~\bibnamefont {Tyng}},\
  }\Doi {10.1103/PhysRevA.66.013602} {\bibfield  {journal} {\bibinfo  {journal}
  {Phys. Rev. A},\ }\textbf {\bibinfo {volume} {66}},\ \bibinfo {pages}
  {013602} (\bibinfo {year} {2002})}\BibitemShut {NoStop}%
\bibitem [{\citenamefont {Theocharis}\ \emph {et~al.}(2006)\citenamefont
  {Theocharis}, \citenamefont {Kevrekidis}, \citenamefont {Frantzeskakis},\
  and\ \citenamefont {Schmelcher}}]{Theocharis}%
  \BibitemOpen
  \bibfield  {author} {\bibinfo {author} {\bibfnamefont {G.}~\bibnamefont
  {Theocharis}}, \bibinfo {author} {\bibfnamefont {P.~G.}\ \bibnamefont
  {Kevrekidis}}, \bibinfo {author} {\bibfnamefont {D.~J.}\ \bibnamefont
  {Frantzeskakis}}, \ and\ \bibinfo {author} {\bibfnamefont {P.}~\bibnamefont
  {Schmelcher}},\ }\Doi {10.1103/PhysRevE.74.056608} {\bibfield  {journal}
  {\bibinfo  {journal} {Phys. Rev. E},\ }\textbf {\bibinfo {volume} {74}},\
  \bibinfo {pages} {056608} (\bibinfo {year} {2006})}\BibitemShut {NoStop}%
\bibitem [{\citenamefont {Pereverzev}\ \emph {et~al.}(1997)\citenamefont
  {Pereverzev}, \citenamefont {Loshak}, \citenamefont {Backhaus}, \citenamefont
  {Davis},\ and\ \citenamefont {Packard}}]{Helium3}%
  \BibitemOpen
  \bibfield  {author} {\bibinfo {author} {\bibfnamefont {S.}~\bibnamefont
  {Pereverzev}}, \bibinfo {author} {\bibfnamefont {A.}~\bibnamefont {Loshak}},
  \bibinfo {author} {\bibfnamefont {S.}~\bibnamefont {Backhaus}}, \bibinfo
  {author} {\bibfnamefont {J.}~\bibnamefont {Davis}}, \ and\ \bibinfo {author}
  {\bibfnamefont {R.}~\bibnamefont {Packard}},\ }\href@noop {} {\bibfield
  {journal} {\bibinfo  {journal} {Nature (London)},\ }\textbf {\bibinfo
  {volume} {388}},\ \bibinfo {pages} {449} (\bibinfo {year}
  {1997})}\BibitemShut {NoStop}%
\bibitem [{\citenamefont {Sukhatme}\ \emph {et~al.}(2001)\citenamefont
  {Sukhatme}, \citenamefont {Mukharsky}, \citenamefont {Chui},\ and\
  \citenamefont {Pearson}}]{Helium4}%
  \BibitemOpen
  \bibfield  {author} {\bibinfo {author} {\bibfnamefont {K.}~\bibnamefont
  {Sukhatme}}, \bibinfo {author} {\bibfnamefont {Y.}~\bibnamefont {Mukharsky}},
  \bibinfo {author} {\bibfnamefont {T.}~\bibnamefont {Chui}}, \ and\ \bibinfo
  {author} {\bibfnamefont {D.}~\bibnamefont {Pearson}},\ }\href@noop {}
  {\bibfield  {journal} {\bibinfo  {journal} {Nature (London)},\ }\textbf
  {\bibinfo {volume} {411}},\ \bibinfo {pages} {280} (\bibinfo {year}
  {2001})}\BibitemShut {NoStop}%
\bibitem [{\citenamefont {Cataliotti}\ \emph {et~al.}(2001)\citenamefont
  {Cataliotti}, \citenamefont {Burger}, \citenamefont {Fort}, \citenamefont
  {Maddaloni}, \citenamefont {Minardi}, \citenamefont {Trombettoni},
  \citenamefont {Smerzi},\ and\ \citenamefont {Inguscio}}]{Cataliotti}%
  \BibitemOpen
  \bibfield  {author} {\bibinfo {author} {\bibfnamefont {F.}~\bibnamefont
  {Cataliotti}}, \bibinfo {author} {\bibfnamefont {S.}~\bibnamefont {Burger}},
  \bibinfo {author} {\bibfnamefont {C.}~\bibnamefont {Fort}}, \bibinfo {author}
  {\bibfnamefont {P.}~\bibnamefont {Maddaloni}}, \bibinfo {author}
  {\bibfnamefont {F.}~\bibnamefont {Minardi}}, \bibinfo {author} {\bibfnamefont
  {A.}~\bibnamefont {Trombettoni}}, \bibinfo {author} {\bibfnamefont
  {A.}~\bibnamefont {Smerzi}}, \ and\ \bibinfo {author} {\bibfnamefont
  {M.}~\bibnamefont {Inguscio}},\ }\href@noop {} {\bibfield  {journal}
  {\bibinfo  {journal} {Science},\ }\textbf {\bibinfo {volume} {293}},\
  \bibinfo {pages} {843} (\bibinfo {year} {2001})}\BibitemShut {NoStop}%
\bibitem [{\citenamefont {Albiez}\ \emph {et~al.}(2005)\citenamefont {Albiez},
  \citenamefont {Gati}, \citenamefont {Folling}, \citenamefont {Hunsmann},
  \citenamefont {Cristiani},\ and\ \citenamefont {Oberthaler}}]{Albiez}%
  \BibitemOpen
  \bibfield  {author} {\bibinfo {author} {\bibfnamefont {M.}~\bibnamefont
  {Albiez}}, \bibinfo {author} {\bibfnamefont {R.}~\bibnamefont {Gati}},
  \bibinfo {author} {\bibfnamefont {J.}~\bibnamefont {Folling}}, \bibinfo
  {author} {\bibfnamefont {S.}~\bibnamefont {Hunsmann}}, \bibinfo {author}
  {\bibfnamefont {M.}~\bibnamefont {Cristiani}}, \ and\ \bibinfo {author}
  {\bibfnamefont {M.}~\bibnamefont {Oberthaler}},\ }\Doi
  {10.1103/PhysRevLett.95.010402} {\bibfield  {journal} {\bibinfo  {journal}
  {Phys. Rev. Lett.},\ }\textbf {\bibinfo {volume} {95}},\ \bibinfo {pages}
  {010402} (\bibinfo {year} {2005})}\BibitemShut {NoStop}%
\bibitem [{\citenamefont {Levy}\ \emph {et~al.}(2007)\citenamefont {Levy},
  \citenamefont {Lahoud}, \citenamefont {Shomroni},\ and\ \citenamefont
  {Steinhauer}}]{Steinhauer}%
  \BibitemOpen
  \bibfield  {author} {\bibinfo {author} {\bibfnamefont {S.}~\bibnamefont
  {Levy}}, \bibinfo {author} {\bibfnamefont {E.}~\bibnamefont {Lahoud}},
  \bibinfo {author} {\bibfnamefont {I.}~\bibnamefont {Shomroni}}, \ and\
  \bibinfo {author} {\bibfnamefont {J.}~\bibnamefont {Steinhauer}},\ }\Doi
  {10.1038/nature06186} {\bibfield  {journal} {\bibinfo  {journal} {Nature
  (London)},\ }\textbf {\bibinfo {volume} {449}},\ \bibinfo {pages} {579}
  (\bibinfo {year} {2007})}\BibitemShut {NoStop}%
\bibitem [{\citenamefont {Williams}\ \emph {et~al.}(1999)\citenamefont
  {Williams}, \citenamefont {Walser}, \citenamefont {Cooper}, \citenamefont
  {Cornell},\ and\ \citenamefont {Holland}}]{Williams}%
  \BibitemOpen
  \bibfield  {author} {\bibinfo {author} {\bibfnamefont {J.}~\bibnamefont
  {Williams}}, \bibinfo {author} {\bibfnamefont {R.}~\bibnamefont {Walser}},
  \bibinfo {author} {\bibfnamefont {J.}~\bibnamefont {Cooper}}, \bibinfo
  {author} {\bibfnamefont {E.}~\bibnamefont {Cornell}}, \ and\ \bibinfo
  {author} {\bibfnamefont {M.}~\bibnamefont {Holland}},\ }\Doi
  {10.1103/PhysRevA.59.R31} {\bibfield  {journal} {\bibinfo  {journal} {Phys.
  Rev. A},\ }\textbf {\bibinfo {volume} {59}},\ \bibinfo {pages} {R31}
  (\bibinfo {year} {1999})}\BibitemShut {NoStop}%
\bibitem [{\citenamefont {Gilmore}\ \emph {et~al.}(1975)\citenamefont
  {Gilmore}, \citenamefont {Bowden},\ and\ \citenamefont {Narducci}}]{Gilmore}%
  \BibitemOpen
  \bibfield  {author} {\bibinfo {author} {\bibfnamefont {R.}~\bibnamefont
  {Gilmore}}, \bibinfo {author} {\bibfnamefont {C.~M.}\ \bibnamefont {Bowden}},
  \ and\ \bibinfo {author} {\bibfnamefont {L.~M.}\ \bibnamefont {Narducci}},\
  }\Doi {10.1103/PhysRevA.12.1019} {\bibfield  {journal} {\bibinfo  {journal}
  {Phys. Rev. A},\ }\textbf {\bibinfo {volume} {12}},\ \bibinfo {pages} {1019}
  (\bibinfo {year} {1975})}\BibitemShut {NoStop}%
\bibitem [{\citenamefont {Lipkin}\ \emph {et~al.}(1965)\citenamefont {Lipkin},
  \citenamefont {Meshkov},\ and\ \citenamefont {Glick}}]{Lipkin}%
  \BibitemOpen
  \bibfield  {author} {\bibinfo {author} {\bibfnamefont {H.~J.}\ \bibnamefont
  {Lipkin}}, \bibinfo {author} {\bibfnamefont {N.}~\bibnamefont {Meshkov}}, \
  and\ \bibinfo {author} {\bibfnamefont {A.~J.}\ \bibnamefont {Glick}},\ }\Doi
  {DOI: 10.1016/0029-5582(65)90862-X} {\bibfield  {journal} {\bibinfo
  {journal} {Nucl. Phys.},\ }\textbf {\bibinfo {volume} {62}},\ \bibinfo
  {pages} {188 } (\bibinfo {year} {1965})}\BibitemShut {NoStop}%
\bibitem [{\citenamefont {Raghavan}\ \emph {et~al.}(1999)\citenamefont
  {Raghavan}, \citenamefont {Smerzi}, \citenamefont {Fantoni},\ and\
  \citenamefont {Shenoy}}]{Raghavan}%
  \BibitemOpen
  \bibfield  {author} {\bibinfo {author} {\bibfnamefont {S.}~\bibnamefont
  {Raghavan}}, \bibinfo {author} {\bibfnamefont {A.}~\bibnamefont {Smerzi}},
  \bibinfo {author} {\bibfnamefont {S.}~\bibnamefont {Fantoni}}, \ and\
  \bibinfo {author} {\bibfnamefont {S.}~\bibnamefont {Shenoy}},\ }\href@noop {}
  {\bibfield  {journal} {\bibinfo  {journal} {Phys. Rev. A},\ }\textbf
  {\bibinfo {volume} {59}},\ \bibinfo {pages} {620} (\bibinfo {year}
  {1999})}\BibitemShut {NoStop}%
\bibitem [{\citenamefont {Leggett}(2001)}]{Leggett}%
  \BibitemOpen
  \bibfield  {author} {\bibinfo {author} {\bibfnamefont {A.~J.}\ \bibnamefont
  {Leggett}},\ }\Doi {10.1103/RevModPhys.73.307} {\bibfield  {journal}
  {\bibinfo  {journal} {Rev. Mod. Phys.},\ }\textbf {\bibinfo {volume} {73}},\
  \bibinfo {pages} {307} (\bibinfo {year} {2001})}\BibitemShut {NoStop}%
\bibitem [{EPA()}]{EPAPS}%
  \BibitemOpen
  \href@noop {} {}\bibinfo {note} {See EPAPS Document No. [number] for
  experimental details}\BibitemShut {NoStop}%
\bibitem [{\citenamefont {Widera}\ \emph {et~al.}(2004)\citenamefont {Widera},
  \citenamefont {Mandel}, \citenamefont {Greiner}, \citenamefont {Kreim},
  \citenamefont {H\"ansch},\ and\ \citenamefont {Bloch}}]{Widera}%
  \BibitemOpen
  \bibfield  {author} {\bibinfo {author} {\bibfnamefont {A.}~\bibnamefont
  {Widera}}, \bibinfo {author} {\bibfnamefont {O.}~\bibnamefont {Mandel}},
  \bibinfo {author} {\bibfnamefont {M.}~\bibnamefont {Greiner}}, \bibinfo
  {author} {\bibfnamefont {S.}~\bibnamefont {Kreim}}, \bibinfo {author}
  {\bibfnamefont {T.~W.}\ \bibnamefont {H\"ansch}}, \ and\ \bibinfo {author}
  {\bibfnamefont {I.}~\bibnamefont {Bloch}},\ }\Doi
  {10.1103/PhysRevLett.92.160406} {\bibfield  {journal} {\bibinfo  {journal}
  {Phys. Rev. Lett.},\ }\textbf {\bibinfo {volume} {92}},\ \bibinfo {pages}
  {160406} (\bibinfo {year} {2004})}\BibitemShut {NoStop}%
\bibitem [{\citenamefont {Erhard}\ \emph {et~al.}(2004)\citenamefont {Erhard},
  \citenamefont {Schmaljohann}, \citenamefont {Kronj\"ager}, \citenamefont
  {Bongs},\ and\ \citenamefont {Sengstock}}]{Sengstock}%
  \BibitemOpen
  \bibfield  {author} {\bibinfo {author} {\bibfnamefont {M.}~\bibnamefont
  {Erhard}}, \bibinfo {author} {\bibfnamefont {H.}~\bibnamefont
  {Schmaljohann}}, \bibinfo {author} {\bibfnamefont {J.}~\bibnamefont
  {Kronj\"ager}}, \bibinfo {author} {\bibfnamefont {K.}~\bibnamefont {Bongs}},
  \ and\ \bibinfo {author} {\bibfnamefont {K.}~\bibnamefont {Sengstock}},\
  }\Doi {10.1103/PhysRevA.69.032705} {\bibfield  {journal} {\bibinfo  {journal}
  {Phys. Rev. A},\ }\textbf {\bibinfo {volume} {69}},\ \bibinfo {pages}
  {032705} (\bibinfo {year} {2004})}\BibitemShut {NoStop}%
\bibitem [{\citenamefont {Gross}\ \emph {et~al.}(2010)\citenamefont {Gross},
  \citenamefont {Zibold}, \citenamefont {Nicklas}, \citenamefont {Esteve},\
  and\ \citenamefont {Oberthaler}}]{Gross}%
  \BibitemOpen
  \bibfield  {author} {\bibinfo {author} {\bibfnamefont {C.}~\bibnamefont
  {Gross}}, \bibinfo {author} {\bibfnamefont {T.}~\bibnamefont {Zibold}},
  \bibinfo {author} {\bibfnamefont {E.}~\bibnamefont {Nicklas}}, \bibinfo
  {author} {\bibfnamefont {J.}~\bibnamefont {Esteve}}, \ and\ \bibinfo {author}
  {\bibfnamefont {M.~K.}\ \bibnamefont {Oberthaler}},\ }\href
  {http://dx.doi.org/10.1038/nature08919} {\bibfield  {journal} {\bibinfo
  {journal} {Nature (London)},\ }\textbf {\bibinfo {volume} {464}},\ \bibinfo
  {pages} {1165} (\bibinfo {year} {2010})}\BibitemShut {NoStop}%
\bibitem [{\citenamefont {Josephson}(1962)}]{Josephson}%
  \BibitemOpen
  \bibfield  {author} {\bibinfo {author} {\bibfnamefont {B.~D.}\ \bibnamefont
  {Josephson}},\ }\Doi {DOI: 10.1016/0031-9163(62)91369-0} {\bibfield
  {journal} {\bibinfo  {journal} {Phys. Lett.},\ }\textbf {\bibinfo {volume}
  {1}},\ \bibinfo {pages} {251 } (\bibinfo {year} {1962})}\BibitemShut
  {NoStop}%
\bibitem [{\citenamefont {Radzihovsky}\ and\ \citenamefont
  {Gurarie}(2010)}]{Radzihovsky}%
  \BibitemOpen
  \bibfield  {author} {\bibinfo {author} {\bibfnamefont {L.}~\bibnamefont
  {Radzihovsky}}\ and\ \bibinfo {author} {\bibfnamefont {V.}~\bibnamefont
  {Gurarie}},\ }\Doi {10.1103/PhysRevA.81.063609} {\bibfield  {journal}
  {\bibinfo  {journal} {Phys. Rev. A},\ }\textbf {\bibinfo {volume} {81}},\
  \bibinfo {pages} {063609} (\bibinfo {year} {2010})}\BibitemShut {NoStop}%
\end{thebibliography}
\end{document}